# A possible relationship between Global Warming and Lightning Activity in India during the period 1998-2009

**Felix Pereira B.[1], Priyadarsini G.[2] and T. E. Girish[3]**

1. Department of Physics, St. Xavier's College, Thumba, Thiruvananthapuram 695586.
2. Department of Physics, Mar Ivanios College, Thiruvananthapuram 695016.
3. Department of Physics, University College, Thiruvananthapuram 695034.

Email: fbpereira@yahoo.co.in

**Abstract**

Lightning activity on a global scale has been studied season wise using satellite data for the period from 1998 to 2009. Lightning activity shows an increasing trend during the period of study which is highly correlated with atmospheric warming. A similar increasing trend of lightning activity is observed in the Indian region during the pre-monsoon season which is correlated with global lightning trends and warming trends of surface temperature in India.

1. Introduction

It is well known that global warming causes changes in the environment. Last decade is known to be the warmest period during the past decades. There are studies on the relationship between atmospheric parameters and thunderstorm activity.[1,2] Studies have proved that lightning activity over the past years have undergone changes.[3] Fluctuations in lightning occurrence depend on a number of factors like season, place, and time.[9]

In this paper we have done a systematic study of lightning activity over the globe and also in the Indian region and its relationship with global warming for the last decade. We have also analysed the season wise changes in lightning activity for the period of study. An increasing trend in lightning activity is observed over the globe for all seasons and a similar behavior is observed in pre-monsoon lightning activity over India. These results are discussed in terms of the possible physical mechanisms which are relevant.

2. Data Analysis

OTD/LIS lightning flash data has been used for our analysis. Monthly average lightning flashes observed globally are calculated for the periods1998 to 2009. Fig.1b shows the variation of lightning activity during the period of study. Sunspot number variation during this period is also shown in Fig.1. It is observed that lightning activity follows the sunspot activity cycle which is minimum during 1999-2000 period and maximum during 2006-2007 periods. The trend of variation of lightning activity is calculated using moving average technique. The earth's northern hemisphere atmospheric temperature anomaly data is used to study global temperature warming. Fig.2 shows the trend of variation of global lightning activity ands northern hemisphere atmospheric temperature anomaly. A significant correlated variation is observed between the two parameters from 1999 to 2009 with a correlation coefficient of 0.8.

The lightning flash data is separated into four seasons namely Winter (Dec, Jan, Feb), Spring (Mach, April, May), Summer (June, July, Aug,), Fall (Sep, Oct, Nov) and the lightning occurrences are plotted in Fig.3. An increasing trend of lightning activity is observed in all seasons during 1999-2009 period. Lightning activity is also observed to be higher during summer in 2003-04 period.

## 3. Lightning observations in the Indian Region

OTD/LIS data is used to study the lightning activity over the Indian peninsular region. The area of the land where lightning flash rate greater than 100/season, is found out from the data for the period 1998-2009. This is taken as the lightning activity index in this region. This is again separated into three seasons: Winter (Dec, Jan, Feb), Pre-monsoon (Mach, April, May) and Monsoon (June, July, Aug, Sep, Oct, Nov). These seasons are selected because in the Indian region, these seasons are the most prominent ones. Fig.3 represents the lightning activity in different seasons over the Indian region during the period of study. An increasing trend of lightning activity variation is observed in the pre-monsoon season. The surface temperature variations during pre-monsoon season for selected west coast stations (Trivandrum and Bombay) also show increasing trends during the period of study (Fig. 5). In the monsoon season, lightning activity is observed to be higher during 2003-04 period in India.

## 4. Discussion

Solar cycle dependence is observed in lightning activity which maximizes during the declining and minimum periods of sunspot activity and minimizes during the sunspot maximum period[5]. Lightning is initiated by cosmic rays[6] and cosmic ray flux in earth's atmosphere is anti-correlated with sunspot cycle.[4] In our analysis an increasing trend of lightning activity is observed during the period of study even though it has solar cycle dependence. Also the correlation between lightning activity and atmospheric temperature anomaly is significant which shows that global warming increases thunderstorm activity. Lightning frequencies in thunderstorms are extremely sensitive to small increase in surface air temperature.[8] A 30% increase in global lightning activity is observed for warmer climate and 24% decrease in global lightning activity is observed for colder climate.[3]

In the season wise study (Fig.3), an increasing trend of lightning activity is observed in all the seasons. In India, an increasing trend of lightning activity is observed during pre-monsoon season. During summer, a prominent increase in lightning activity is observed from 2002 to 2005, during which period, solar energetic particle events are more frequent.[7] Lightning activity in India can be modulated by sunspot cycle and atmospheric warming, which can initiate convective phenomena like thunderstorm activity. Surface temperature in the Indian region also showed an increasing trend (Fig. 5). An increasing trend of lightning activity is observed in the pre-monsoon season in India, which is highly correlated with global lightning trends and surface temperature variations observed during the same period. On a global scale, the relationship between precipitation ice water path and lightning flash density is relatively invariant between lands, ocean and coastal regimes.[8]

**5. Conclusions**

Satellite observations of global lightning activity during 1998–2009 period show an increasing trend for all seasons during which period, severe global warming trends are observed. During the pre-monsoon season, lightning activity in India show a similar increasing trend with global lightning activity and warming trends of surface temperature over Indian region during the same season.

**Acknowledgement**

We thank the NASA LIS/OTD Science team for providing us the data.

**Figure Captions**

Fig. 1: Monthly average Sunspot number (*Top*) and Monthly occurrence of Lightning flashes (*Bottom*) for the period 1998 – 2009.

Fig.2: Variation of Lightning flashes/year (*Thin line*) and Earth's Northern Hemisphere atmospheric Temperature anomaly (*Thick line*) for the period 1998 – 2009.

Fig. 3: Variation of Occurrence of Lightning flashes per season during winter, spring, summer and fall seasons during 1998 -2009 period.

Fig. 4: Variation of area of land where occurrence of lightning flashes greater than 100 for a) pre monsoon b) monsoon c) winter and d) independent of seasons, for the period 1998- 2009.

Fig. 5: Pre-monsoon surface temperature variations observed in Trivandrum (*Top*) and Bombay (*Bottom*).

Figure 1

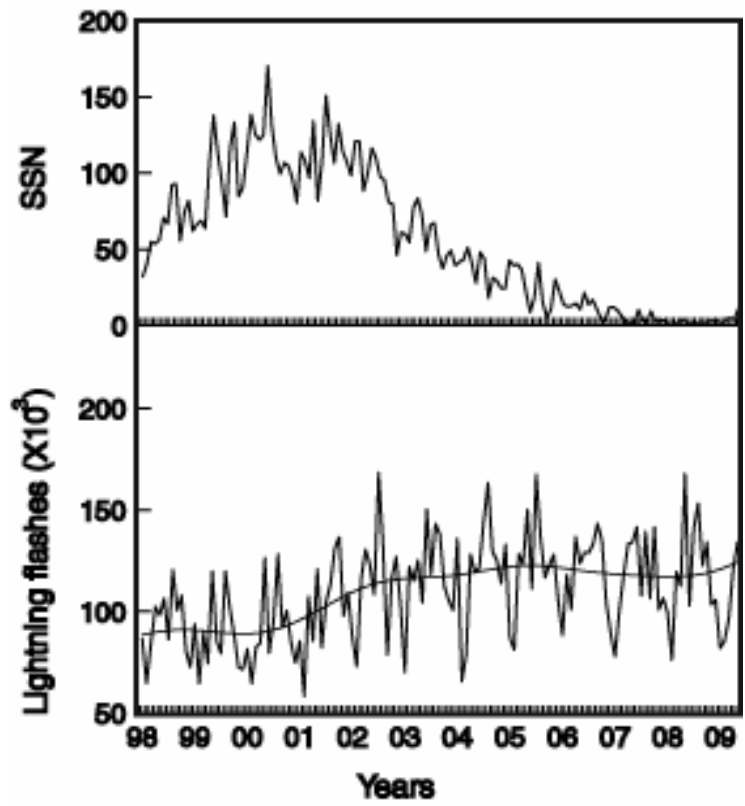

Figure 2

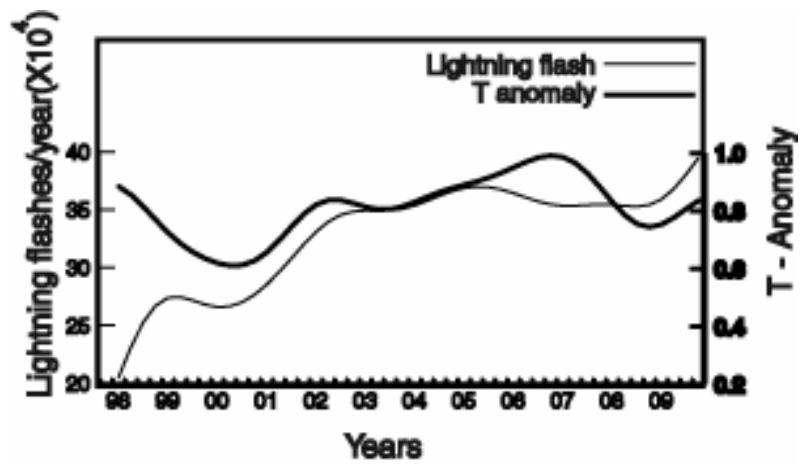

Figure 3

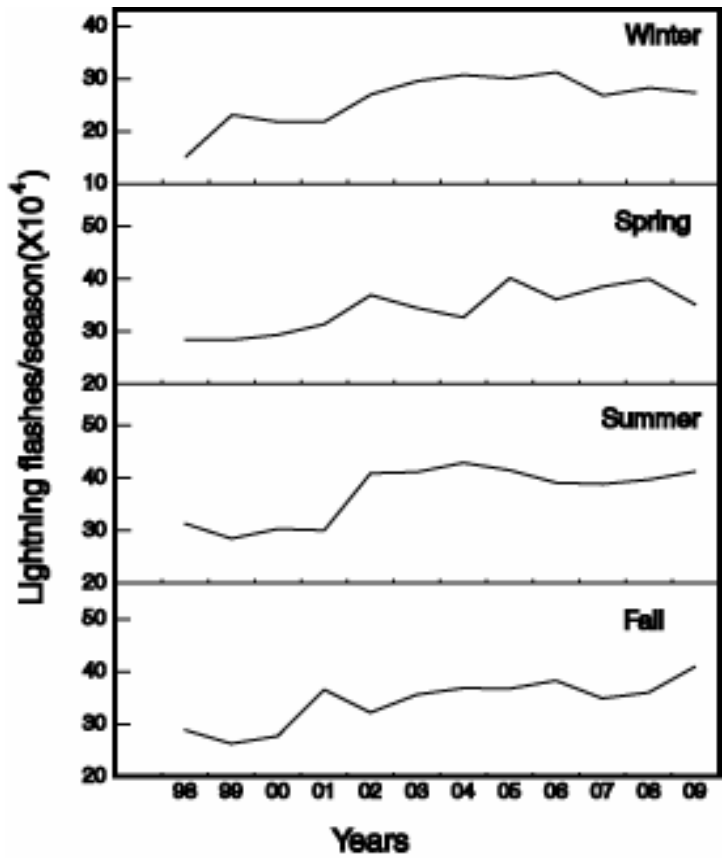

Figure 4

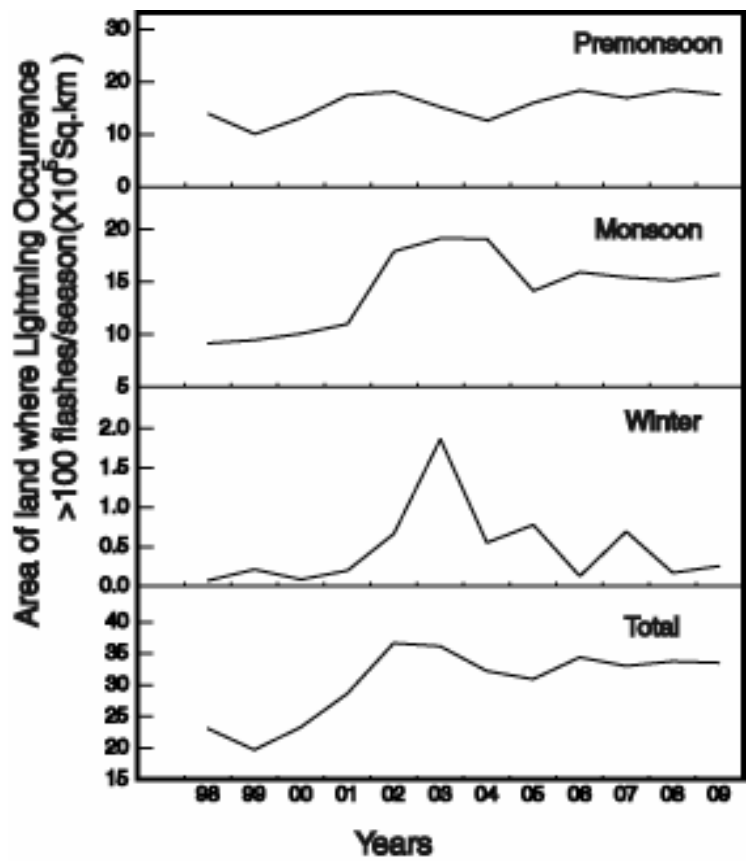

Figure 5

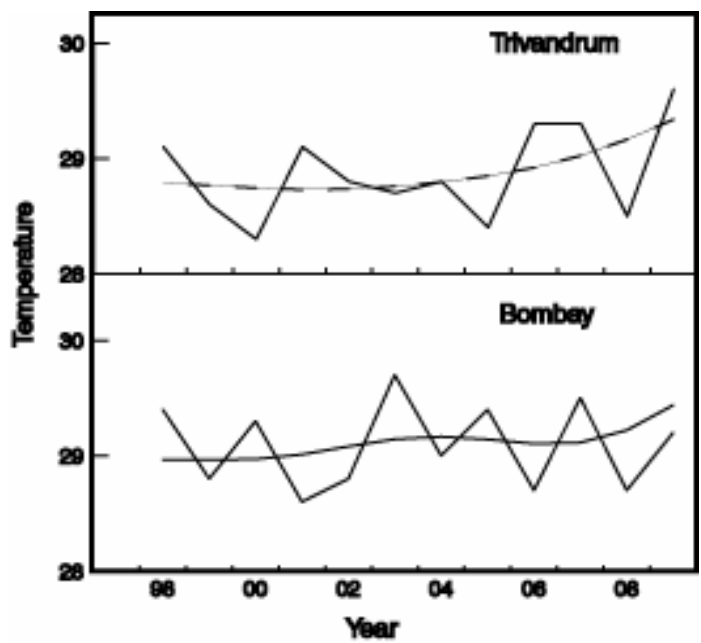